\documentclass[twoside, onecolumn, english, aps, prl, notitlepage, preprintnumbers]{revtex4-1}
\usepackage{bbm, bm}
\usepackage[dvips]{graphicx}
\usepackage[centertags]{amsmath}
\usepackage{amssymb}
\usepackage{amsfonts}
\usepackage{psfrag}
\usepackage{subfigure}
\def\exp{\text{Exp\,}}

\def\probP{{\cal P}}

\begin{document}
\title{The role of multiple marks in epigenetic silencing and the emergence of a stable bivalent chromatin state}
\author{Swagatam Mukhopadhyay}\affiliation{Cold Spring Harbor Laboratory, Cold Spring Harbor, NY, USA}
\author{Anirvan M. Sengupta}\affiliation{BioMaPS and Department of Physics, Rutgers University, Piscataway, NJ, USA}
\date{\today}

\begin{abstract}
We introduce and analyze a minimal model of epigenetic silencing in budding yeast, built upon known biomolecular interactions in the system. Doing so, we identify the epigenetic marks essential for the bistability of epigenetic states. The model explicitly incorporates two key chromatin marks, namely H4K16 acetylation and H3K79 methylation, and explores whether the presence of multiple marks lead to a qualitatively different systems behavior. We find that having both modifications is important for the robustness of epigenetic silencing. Besides the silenced and transcriptionally active fate of chromatin, our model leads to a novel state with bivalent (i.e., both active and silencing) marks under certain perturbations (knock-out mutations, inhibition or enhancement of enzymatic activity). The bivalent state appears under several perturbations and is shown to result in patchy silencing. We also show that the titration effect, owing to a limited supply of silencing proteins, can result in counter-intuitive 
responses. The design principles of the silencing system is systematically investigated and disparate experimental observations are assessed within a single theoretical framework.  Specifically, we discuss the behavior of Sir protein recruitment, spreading and stability of silenced regions in commonly-studied mutants (e.g., \emph{sas2}$\varDelta$, \emph{dot1}$\varDelta$) illuminating the controversial role of Dot1 in the systems biology of yeast silencing. 
\end{abstract} 
\maketitle 
\section*{Author Summary}

Epigenetics is the study of heritable phenotypic variations that are not caused by changes in the genotype. Silent Information Regulator (SIR) silencing in budding yeast is an important model system for epigenetics. The standard model of silencing relies on feedback, mediated by chromatin modifications (for example, deacetylation of histone residues) which lead to enhanced recruitment of chromatin modifiers. However, the SIR mechanism is not completely understood and it is important to investigate whether as-yet-undiscovered components alter the systems design in a fundamental way. We address this question using minimal models constructed from experimentally known interactions. Rather than building a detailed network model with parameters to fit for quantitative predictions, we build an effective model and study its bifurcation diagram which leads to robust qualitative predictions on the nature of mutants. This minimal modeling delineates a phase space with qualitatively different epigenetic mechanisms and 
states; some of which arise from drug/genetic perturbations and exhibit large cell-to-cell variation in chromatin marks. Our methodology can be applied to the study of epigenetic chromatin silencing in other model systems, especially Polycomb silencing, and reveals engineering principles that may be of broad relevance.  

\section*{Introduction}

Understanding the design principles of epigenetic silencing phenomena poses challenges to experimentalists and theorists alike, primarily owing to the complexity of interactions between epigenetic marks discovered in recent years~\cite{Turner2008, Kouzarides2007, Latham2007}. In the model organism \emph{Saccharomyces cerevesiae}, since the discovery of telomeric position effect~\cite{Gottschling1990, Aparicio1991}, intense experimental activity has continued to unravel the many facets of epigenetic silencing in telomeric, Hidden Mating (HML/HMR) loci and ribosomal-DNA (rDNA) regions. However, a complete picture of how epigenetic fate is established, maintained and inherited faithfully is still lacking. 

Two universal characteristics of epigenetic phenomena are the switch-like behavior in the expression of genes proximal to silenced chromatin domains and the robust reestablishment of silenced domains through multiple cell cycles. Both of these phenomena impose constraints on any mathematical model of silencing. The nucleation and spreading of silencing in budding yeast, that strongly argues for the bistability of the mechanism, is well-documented by now~\cite{Rusche2002, Rusche2003, Talbert2006}, though the mechanism of establishment, maintenance and inheritance is still a matter of active research~\cite{Katan-Khaykovich2005, Martins-Taylor2004, Dion2007, Probst2009, Osborne2009, Kaufman2010, Radman-Livaja2011}. At least two different mechanisms of silencing spreading have been proposed in the literature. We call the first mechanism the \emph{polymerization model} of spreading~\cite{Moazed1997, Moazed2004,Rudner2005, Onishi2007}, whereby interactions of proteins amongst themselves lead to a spread of 
silencing along the chromatin. We call the second mechanism the \emph{histone modification feedback mechanism}, whereby the recruitment and spreading of silencers is controlled by histone modifications and silencing/transcriptional feedbacks; various anti-silencing marks etc. block entire regions of chromatin from being silenced. Are these two mechanisms redundant? We explore these mechanisms and argue that they play synergistic role in the robustness of the system; an important conceptual outcome of our approach. 

Multiple chromatin modifications (marks) are thought to be responsible in carrying epigenetic information~\cite{Berger2007, Kouzarides2007,Latham2007}. Anti-silencing histone marks, primarily acetylation and methylation, are known to be key players in budding yeast silencing~\cite{Kurdistani2003, Dhillon2002,Ng2002a}. What role does multiple histone marks play in the dynamics and stability of the silencing phenomena? Why are multiple anti-silencing marks needed? Despite recent efforts at mathematical modeling of epigenetic chromatin modification in budding yeast~\cite{Sedighi2007, David-Rus2009, Kelemen2010} and in other systems~\cite{Dodd2007, Micheelsen2010, Angel2011}, theoretical understanding on the role of multiple marks in epigenetic bistability remain incomplete. Moreover, currently there is a gap in the literature on mathematical models that explain epigenetic bistability in terms of \emph{local} biochemical interactions. Building informative models is not necessarily impaired by our lack of 
knowledge of the biochemistry or reaction rates, and non-trivial qualitative predictions can be obtained from rather minimal models providing an integrative understanding of experiments in the field. We illustrate the role of multiple marks in epigenetic bistability and robustness of silenced/active regions using a minimal model. 

Experiments largely rely on the behavior of knock-out mutants to unravel the components of the yeast silencing system. Less well-studied are the inhibition/enhancement (of enzymatic activity) or over/under-expression (of proteins), though overexpression of Sir proteins and Dot1 have been examined~\cite{Renauld1993, Maillet1996, Singer1998, Pryde1999, Imai2000a, Ng2003a, Osborne2011}. We argue that compared to knock-out mutants, inhibition etc. is better suited at revealing the engineering design principles of the system. Using theory and simulations we infer the pivotal role of Dot1 in establishing stable silencing domains and resolve seemingly contradictory experimental observations on this matter. We also argue that as long as the set of known interactions included in our model are retained, our overall conclusions are robust to the addition of more complex interactions, either known or yet to be discovered. 

\subsection*{Primer on molecular biology of budding yeast silencing}  
\label{sec:PrimerMolBio}

To provide the relevant context of our approach to a wider audience, we briefly review the molecular biology of yeast silencing system and its central puzzles. In budding yeast, the \underline{S}ilent mating-type \underline{I}nformation \underline{R}egulator (SIR) proteins have been identified to play a pivotal role in heterochromatin formation~\cite{Renauld1993, Hecht1995,Strahl-Bolsinger1997,Allisbook2007}. The Sir complex includes four proteins, of which Sir1 facilitates assembly of Sir2-4 at the \underline{H}idden \underline{M}ating (HML/HMR) Cassettes, in concert with proteins factors like Orc, Rap1 an Abf1, all of which have specific binding sites at the silencer of HML/HMR. In telomeres, the players are slightly different; Sir1 is not needed for the nucleation of Sir complex~\cite{Strahl-Bolsinger1997, Allisbook2007} and yKu DNA-end binding complexes is important. One of the proteins of the Sir complex, Sir2, is a NAD-dependent deacetylase~\cite{Tanner2000}. The acetylation of Histone tails 
antagonizes silencing~\cite{Johnson1992, Hoppe2002, Carmen2002, Kurdistani2003, Kristjuhan2003}. The deacetylase activity of Sir2 is critical for the spread of silencing by presumably increasing Sir3-Sir4 affinity for nucleosomes~\cite{Martino2009}. Specifically, acetylation of H4K16 by Sas2 (\underline{S}omething \underline{A}bout \underline{S}ilencing) activity undermined the spread of silencing~\cite{Rusche2002, Suka2002, Rusche2003, Kimura2002}. Recent studies have established that besides the interaction of Sir proteins with nucleosomes~\cite{Sampath2009}, interactions between Sir proteins themselves are crucial in the spread of silencing~\cite{Chang2003, Rudner2005, Martino2009}. This spreading in telomeres appears to be unhindered by silencer boundary element and is arguably stochastic~\cite{Pryde1999, Kristjuhan2003,Martin2004, Millar2006, Taddei2009}. In contrast, in the HML/HMR loci boundary elements prevent the spread of Sir complex~\cite{Bi2002, Rusche2003}. Sir spreading happens in only a few of 
the natural telomeres~\cite{Pryde1999, Radman-Livaja2011, Takahashi2011}. Moreover, mechanisms like the clustering of telomeres anchored to the nuclear envelope is perhaps important in subtelomeric silencing~\cite{Taddei2009}. 

Sir2 does not exclusively deacetylate H4K16 residue. Other residues like H3K9 and H3K14 are also deacetylated moderately~\cite{Imai2000a}, and H4K56 extensively by Sir2~\cite{Xu2007}. However, the spreading of Sir2 compete primarily with H4K16Ac~\cite{Suka2002}. The compaction into silent heterochromatin requires further deacetylation of H4K56Ac~\cite{Xu2007}. The effect of spread of Sir proteins on silencing is subtle---spreading is necessary but not sufficient for heterochromatin formation and Sir binding is broader in scope~\cite{Radman-Livaja2011}. Though the compaction of chromatin owing to stable Sir-complex-binding blocks association of RNA polymerase II thereby preventing transcriptional activity~\cite{Rusche2003}, Sir binding and spreading itself may be a very dynamic phenomenon. Not surprisingly, Sir protein binds at various loci which are not necessarily silenced, as deduced from Chromatin Immunoprecipitation (ChIP) studies~\cite{Lieb2001,Radman-Livaja2011}. The association of Sir proteins does 
not immediately lead to gene silencing~\cite{Kirchmaier2006} arguing for added steps to the formation of heterochromatin~\cite{Lau2002, Martins-Taylor2004, Katan-Khaykovich2005, Probst2009}. Nevertheless, Sir protein association is a prerequisite in establishing an inheritable pattern of silencing in the telomeric and HML/HMR loci.  

Active sites in budding yeast chromatin are not only hyper-acetylated at particular histone tail sites but also tend to be hyper-methylated at certain key residues~\cite{Singer1998, Leeuwen2002a, Ng2002a, Ng2003a, Millar2006}. One of the DOT (\underline{D}isruptor \underline{O}f \underline{T}elomeric Silencing) proteins, Dot1, methylates H3K79 residue and competes in binding with Sir3 for the same basic patch on the histone core region (H3K79)~\cite{Leeuwen2002a, Altaf2007, Sawada2004, Martino2009}. Dot1 can mono- di- or tri- methylate this residue distributively~\cite{Frederiks2008}. Studies have revealed that H4K16Ac displaces Sir3 binding, thereby aiding Dot1-mediated methylation of H3K79~\cite{Altaf2007}. What other histone marks (or other factors) regulate transcriptional activity is a complex question~\cite{Saunders2006}. The enzymatic activity of Dot1 itself is modulated by various other factors. For example, Paf1 complex (RNA \underline{P}olymerase \underline{A}ssociating \underline{F}actor) which is 
known to be an important factor in transcription elongation plays a crucial role in Dot1 methylase activity~\cite{Wood2003, Krogan2003a, Zhang2009}. In fact, such \emph{positive feedback} from transcriptional elongation into establishing transcriptionally-active marks is not uncommon. For example, H3K4 methylation~\cite{Santos-Rosa2002, Venkatasubrahmanyam2007, Yang2008}, which is an active mark in yeast, requires Paf1~\cite{Zhang2009} for being methylated by Set1~\cite{Roguev2001}.

The phenomenon of yeast silencing has several other components many of which are poorly understood. Though the structure and interaction domains of Sir proteins are well-studied~\cite{Gasser2001, Chang2003, King2006, Martino2009, Sampath2009, Norris2010}, protein factors, histone modifications etc. which interact with the system continue to be uncovered. For example, Rad6 dependent ubiquitylation of H2B-K123 influences methylation by Dot1 and Set1~\cite{Shahbazian2005, Yang2008}. Ubiquitylation is a histone mark implicated in transcriptional initiation and elongation~\cite{Weake2008}, and such a \emph{trans-histone} regulatory pathway acts as a feedback into the establishment of transcriptional activity. Another such feedback in establishment of active marks is the interaction of Dot1 with a histone acetyltransferase~\cite{Stulemeijer2011}. Sumoylation is a silencing histone mark, and might play an important role in the silencing of subtelomeric regions~\cite{Nathan2006, Fukuda2006}. Variants of the 
histones~\cite{Kamakaka2005, Sarma2005}, particularly H2A.Z, plays a complex role in transcriptional activity, and has been shown to deter the ectopic spread of silencing~\cite{Venkatasubrahmanyam2007}. The topic of histone variants invite a host of connections between chromatin assembly and transcriptional activity. These influences are perhaps peripheral to the central mechanism resolving epigenetic fates~\cite{Enomoto1998a, Kirchmaier2001, Osada2001}. 

We argue that the epigenetic fates are reinforced both from \emph{transcriptional} and \emph{silencing feedbacks}---our focus is the core mechanism, summarized in Fig.~\ref{fig:cartoon}. 
Recent experiments~\cite{Radman-Livaja2011} have shown that binding of Sir3 protein is not indicative of heterochromatin and many genes in euchromatin, including highly transcribed genes, may show wide-spread Sir3 binding. Such a co-occurrence of silencing and active marks called bivalent chromatin states has been discussed in the context of Polycomb silencing~\cite{Schwartz2008} and that of HP1 binding in expressed exons~\cite{Johansson2007}. We show how stable bivalent states emerge naturally on modelling the known interactions. In the next section we define the mathematical model based on this \emph{minimal} picture.  

\section*{Results}
\subsection*{The model and its phase space} 

In this section we introduce the key components of our model in order to facilitate discussion, further details are relegated to Materials and Methods and Text S1. Continuing on the spirit of earlier work involving one of the authors~\cite{Sedighi2007}, we define several local states that represent the various modifications of nucleosomes.  Five possible nucleosome-states are considered at nucleosome $i$: $S_i$, $A_i$, $M_i$, $E_i$ and $U_i$. The $S$ state is Sir-complex-bound (without making a distinction of the various Sir proteins). The $A$ state is acetylated histone (H4K16Ac). Multiple histone tails on the same nucleosome can get acetylated---we model the average level of acetylation. The $M$ is the methylated state (H3K79Me), and we treat the multiple levels of methylations on the average. The $U$ is unmodified state--- it is neither acetylated, methylated or Sir-protein bound. The $E$ state is the transcriptionally active state and has multiple histone marks, particularly both methylation and 
acetylation marks. 

The model considers the following biochemical processes, see Fig.~\ref{fig:cartoon-2} for a pictorial representation of the model. Index $j$ are for nucleosomes within a local neighborhood of nucleosome $i$: 
\begin{itemize} 
\item{{\bf{Basal Sir complex binding}} at rate $\rho_0$: $U_i \stackrel{\rho_0}{\rightarrow} S_i.$}
\item{{\bf{Basal Dot1-mediated methylation}} at rate $\beta_0$:  $U_i \stackrel{\beta_0}{\rightarrow} M_i, A_i \stackrel{\beta_0}{\rightarrow} E_i.$}
\item{{\bf{Basal Sas2-mediated acetylation}} at rate $\alpha$:  $U_i \stackrel{\alpha}{\rightarrow} A_i, M_i \stackrel{\alpha}{\rightarrow} E_i.$ }
\item{{\bf{Cooperative deacetylation by Sir2}} at rate $\Gamma$:  $A_i S_j \stackrel{\Gamma}{\rightarrow} U_i S_j$, where $j$ is a neighboring nucleosome to $i$.}
\item{{\bf{Cooperative methylation}} at rate $\beta$: `transcriptional feedback': $U_i E_j\stackrel{\beta}{\rightarrow} M_i E_j, A_i E_j \stackrel{\beta}{\rightarrow} E_i E_j$.}
\item{{\bf{Cooperative Sir binding}} rate $\rho$: `Silencing polymerization': $U_i S_j \stackrel{\rho}{\rightarrow} S_i S_j$. }
\item{Global and local rates of loss of marks $\eta$ that converts all other states to $U$.}
\end{itemize} 

The mathematical details of the model are relegated to Methods and Materials and Text S1. Therein we establish that the key parameters are the rates of Sas2 activity $\alpha$, the cooperative Dot1 activity $\beta$, the cooperative Sir2 activity $\Gamma$ and the cooperative Sir binding $\rho$. The first result of our model is the \emph{phase space}---a four dimensional parameter space with each point, being a distinct choice of parameters, produces solutions for the density of marks in the model. The distinct properties of such solutions, like stability and density of marks, define distinct phases. For example, regions of \emph{bistability} are where stable active ($E$) and silent ($S$) states can coexist. \emph{Stability} in this context is against the loss of epigenetic marks through various perturbations---histone turnover, DNA replication, stochasticity in enzymatic reactions, deacetylase/demethylase activity etc. and maintaining heritable distinct epigenetic fates for the chromatin loci. \emph{Bivalent} region is where the stable solution exhibits simultaneously high local density of active and silenced marks. 

The ranges of values of the parameters we introduce above are unknown. Various loci in the wild type cell corresponds to different combinations of parameters and therefore maps to distinct points in phase space. These wild-type points should lie in regions where perturbations are unlikely to affect the epigenetic fates of the loci---a requirement of robust engineering design. We have studied aspects of this robustness elsewhere~\cite{David-Rus2009, Mukhopadhyay2010a}, wherein we have discussed the requirements on engineering design for faithful reestablishment of epigenetic states from redistribution and dilution of modified histones during mitosis. A quantitative measure of robustness of a epigenetic states is the volume and shape of parameter space where the state is supported~\cite{Dayarian2009}. 

We first outline the phase space. The model has four types of uniform steady-state solutions (in \emph{mean-field} analysis, see Materials and Methods and Text S1): the \emph{silenced} state which has probability weight predominantly in the state $S$, the \emph{active} state with weight predominantly in the state $E$, the \emph{intermediate} state with weight distributed in $U$, $A$ and $M$, and the \emph{bivalent} state with weight distributed predominantly over $S$ and $E$. The model exhibits bistability between the \emph{silenced} state and the \emph{intermediate} state, and the \emph{silenced} state and the \emph{active} state. The \emph{bivalent} state is \emph{monostable} and is not to be confused with the \emph{bistable} regions.  A novel outcomes of our analysis is that only four distinct \emph{stable} epigenetic fates emerge in all of phase space. 

Visualizing the four dimensional phase space is challenging---we present different two dimensional sections of the phase space for the purpose. Each section is defined by fixing the values of any two of the four parameters. The section shown in Fig.~\ref{fig:firstcut}, for which $\Gamma$ and $\rho$ were fixed, is a representational one for the model's phase space. Only stable solutions are shown. The region of stable \emph{silenced} state is depicted by red diamonds, the stable \emph{active} with green stars, stable \emph{bivalent} with magenta circles, and stable \emph{intermediate} with blue crosses in our phase plots. We adhere to this convention throughout the paper. 

\subsection*{The Sir titration effect results in a local rate $\rho$ of Sir recruitment} 

The supply of Sir proteins is limited in order to prevent ectopic silencing in wild type.  The reaction rates $\rho_0$ and $\rho$, for basal and cooperative Sir binding respectively, are proportional to the concentration of ambient Sir proteins. In telomeres where no obvious boundary elements are present to limit the spread of silencing proteins, the concentration of ambient Sir proteins self-adjusts such that the spreading of silencing is stochastically stationary; the choice of parameters for which such stationarity is achieved is known as the \emph{zero velocity line}~\cite{Sedighi2007}. Away from this line, but still within a bistable region, either the \emph{active} state or the \emph{silenced} state spreads. 

Denoting by $S$ the the total number of Sir proteins, and $s$ the density of Sir proteins, the equation that determines $\rho$ is 
\begin{equation}
\begin{split} 
S_{\text{total}} &= S_{\text{bound}} + S_{\text{ambient}}\\
\rho \left( s_{\text{ambient}} \right)  &= \rho\,\frac{S_\text{total} - S_\text{bound}}{V}, 
 \end{split} 
 \label{eq:titration}
\end{equation}
where $V$ is the volume of the cell nucleus. As silencing spreads, $S_{\text{bound}}$ increases, and $\rho$ decreases. Such a self-adjusting $\rho$ is the titration effect on Sir spreading. Sir spreading needs to happen only in a few of all the telomeres~\cite{Radman-Livaja2011} in order for this effect to be observed and be important for perturbations. The zero-velocity line determines how the local Sir recruitment rate evolves under perturbations, as ellaborated in the next sections. 

\subsection*{Sir2 inhibition leads to bivalent state}

The epigenetically stable bivalent state is a novel outcome of our model, and we discuss the implications of its emergence under the inhibition of Sir2 deacetylase activity, i.e., on reducing $\Gamma$ in our model. Because of the titration effect, whenever a single parameter is changed the parameter $\rho$ self-adjusts to maintain the system on the zero-velocity line. In effect, the silencing front progresses or retreats to be stochastically stationary again. In Fig.~\ref{fig:zero-velo-Sir2}, we show the cross section of the phase space for fixed Dot1 activity ($\beta$) and Sas2 activity ($\alpha$). The zero velocity line is determined in stochastic simulations, as described in Materials and Methods, and is plotted over the mean-field phase diagram. As we decrease Sir2 deacetylase activity ($\Gamma$), $\rho$ traces the zero-velocity line driving the system eventually out of the bistable region (wild-type) and into the region of monostable bivalent state, see Fig.~\ref{fig:zero-velo-Sir2}. 
 
The implications of this observation are two fold---
\begin{itemize}
\item{Inhibition of Sir2 activity eventually drives the wild-type bistable system, where silencing and active states are stable, into a monostable region where bivalent states are stable.}
\item{The inhibition of Sir2 activity leads to an excess of ambient Sir proteins.}
\end{itemize} 
The first observation is a non-trivial prediction of the model. Though there has been reports of defective silencing boundary~\cite{Suka2001, Suka2002, Garcia2002, Kimura2002} in telomeres for Sir2 and Sas2 perturbations, the nature of the defect for Sir2 inhibition has not been made precise. 

The second observation implies that effective Sir cooperativity ($\rho$) increases locally for decreasing deacetylase activity ($\Gamma$) of Sir2, and is relevant to the spatial feature of the \emph{bivalent} state, which we investigate in lattice simulations. We study the spreading and steady-state occupancy of silencing marks with a nucleating center for Sir binding at one lattice end. We observe that the bivalent state is stochastically established Sir occupancy. The silencing boundary is ill-defined as shown in Fig.~\ref{fig:Sas2MutantSpreading}.  A precise characterization of the typical size of these patches is the \emph{correlation length} of Sir occupancy---a measure of how influential the state of a nucleosome is on the state of neighboring ones, see Fig.~\ref{fig:corr-length1}. We measure the correlation length of $S$ marks for the system self-tuning $\rho$ along the zero-velocity line shown in Fig.~\ref{fig:zero-velo-Sir2}. The correlation length is high, as expected, in the bistable region where 
silencing domains are established. However, the bivalent state can maintain rather high correlation lengths resulting in local patches of silencing domains; see further discussion in the next section. The bivalent state is truly \emph{bivalent}, in the sense that nearby nucleosomes carry opposite marks, only in selected regions of the parameter space.

\subsection*{Role of Dot1 in the systems design} 
\label{sec:CuriousDot1}

The precise role of Dot1 is the yeast silencing continues to be active debated~\cite{Takahashi2011, Osborne2009, Osborne2011, Stulemeijer2011, Yang2008}, and a commonly used assay to report the telomeric position effect variegation (TPEV) has been brought to question for \emph{dot1$\varDelta$} strain~\cite{Rossmann2011}. Experiments~\cite{Ng2002a, Ng2003a, Leeuwen2002a, Welsem2008, Takahashi2011} have focused on both \emph{dot1$\varDelta$} and overexpression of Dot1, but not inhibition. In this section, we summarize the bearings of our model on Dot1 perturbations. 

The \emph{dot1$\varDelta$} strain is bistable in our model, but the bistability is for \emph{intermediate} and \emph{silenced} states, as seen in Fig.~\ref{fig:rhobetacut2} to be the region of overlap of \emph{silenced}-stable (red diamonds) and \emph{intermediate}-stable (blue crosses). The bistability achieved in the absence of Dot1 activity may cast doubt on the necessity of Dot1 in our minimal model construction. We argue that Dot1 plays critical role in establishment of heritable silenced domains. 

We argue that measuring Sir occupancy, as opposed to transcriptional activity resolves the pivotal role played by Dot1. This is because transcriptional activity does not imply a unique state of local histone modifications. Though hyper-acetylation and hyper-methylation is associated with transcriptional activation in genome-wide studies~\cite{Schuebeler2004}, methylation is not required for moderate transcriptional activity~\cite{Katan-Khaykovich2005, Yang2008, Jin2009}. A subtlety of our model, and a potential criticism, is our definition of an \emph{active} state, viz., states having both acetylation and methylation marks as being too stringent. However, we can always consider $A$ domains to be moderately transcribed. We observe in Fig.~\ref{fig:rhobetacut2} that there is a critical value of $\beta$, which is given by $ \beta_\text{crit}  = \frac{(1 + \alpha)^2}{\alpha(2+\alpha)}$ (see Text S1), above which the system is \emph{active-silenced} bistable and below which the the system is \emph{intermediate-silenced} 
bistable. In the presence of non-zero basal rates of Dot1 activity ($\beta_0$) and Sir binding ($\rho_0$) the sharp transition between the two bistable regions becomes a crossover (see Text S1). Therefore, if we use transcriptional activity as a measurement, the model implies that increasing Dot1 level ($beta$) from null simply makes transcription robust. Is Dot1 redundant in epigenetic bistability? We now argue that a clearer picture emerges on using Sir occupancy as a measurement.  We drive the system in simulation through the \emph{silenced}-\emph{active}-bistable (wild-type) to the \emph{silenced}-\emph{intermediate}-bistable region by reducing $\beta$, thereby recreating the effects of Dot1 inhibition. We also drive the system from the \emph{silenced}-\emph{active}-bistable (wild-type) bistable (wild type) to \emph{bivalent}-monostable region thereby recreating the effects of Dot1 overexpression. Key observations from the model are summarized in the subsections below. 

\subsubsection*{Inhibition of Dot1 leads to defective region boundaries} 

Dot1 inhibition ($\beta < \beta_\text{crit}$) leads to lower local density of Sir occupancy in telomeres but higher fraction of Sir proteins to be chromatin-bound, leading to Sir depletion and ill-defined silenced domains, see Fig.~\ref{fig:Dot1inhibitionSir} and Fig.~\ref{fig:Dot1boundaries}. In our model, Dot1 inhibition leads to the system crossing the \emph{silenced}-\emph{active}-bistable (\emph{wild type}) to the \emph{silenced}-\emph{intermediate}-bistable region.  In model simulations, we show that the wild type system has higher density of Sir occupancy---Dot1 inhibition leads to lowering of local Sir occupancy in the \emph{silenced}-\emph{intermediate}-bistable region, see Fig.~\ref{fig:Dot1inhibitionSir}.  However, the fraction chromatin-bound Sir exhibits the opposite trend with increased fraction for Dot1 inhibition. More chromatin-bound Sir leads to depletion of ambient Sir proteins. The consequence of Sir depletion in the \emph{dot1$\varDelta$} strain by the telomeres can lead to reduced 
silencing in hidden mating loci~\cite{Welsem2008, Yang2008}. Strikingly, the lower density of Sir occupancy in \emph{dot1$\varDelta$} has recently been observed in experiments~\cite{Takahashi2011} in agreement with our result.

In the \emph{dot1$\varDelta$} strain, the silencing regions are variegated and the domains are ill-defined, see Fig.~\ref{fig:Dot1boundaries}. Methylation---and multiple histone marks in general---play a key role in establishing and maintaining heritable silenced and active domains distinguished by sharp boundaries. The effect of tuning Dot1 activity from null is presented in Fig.~\ref{fig:rhobetacut2}, corresponding to Fig.~\ref{fig:zero-velo-Sir2}. Recall that the self adjustment of $\rho(s_\text{ambient})$ owing to titration effect maintains the steady-state system on the zero-velocity line. A sample of the relevant cuts through the phase diagram is also shown in Fig.~\ref{fig:rhobetacut2}.

In order to characterize both the bistable regions further, we measure the correlation length of the $S$ marks as the system travels along the zero-velocity line; see Fig.~\ref{fig:corr-length2} and refer to Fig~\ref{fig:rhobetacut2}. For clarity of the discussion below, we quote numerical values of $\beta$ specific to the section of phase space presented. Ovbiously, these values are in general not meaningful. With that caveat---the range $\beta \in [0,8]$ in Fig.~\ref{fig:corr-length2} and Fig.~\ref{fig:rhobetacut2} is relevant for the present discussion, where the crossover between the two bistable regions is approximately at $\beta~1.5$. In the wild type ($1.5 \lesssim \beta \lesssim 9 $) silenced state is stable and we expect strong correlations in Sir occupancy, which should survive Dot1 inhibition (lowering $\beta$) even after active state loses stability for $\beta \lesssim 1.5$). Accordingly, we observe in Fig.~\ref{fig:corr-length2} that the correlation length drops but does retain moderate values 
for strong Dot1 inhibition. The drop is explained by the defective establishment of the silencing domain in telomeres, see Fig.~\ref{fig:Dot1boundaries}. 

The survival of correlations indicates that Dot1 perturbation does not necessarily eliminate silencing in shorter loci like HML/HMR with boudary elements that may nucleate silencing, provided their length is comparable to the correlation length of Sir occupancy~\cite{Leeuwen2002a, Osborne2011, Rossmann2011, Takahashi2011}. We cannot determine this correlation length without further knowledge of the wild-type parameters and other details of the chromatin polymer. Nevertheless, we make a strong case for reporting the local level of Sir occupancy in the loci, as opposed to reporting transcription to study the role of Dot1. 

\subsubsection*{Over-expression of Dot1 leads to patchy silencing and bivalent state} 

Over-expression of Dot1 eventually pushes the system to the bivalent state, which is stable, and therefore, potentially heritable. We observe that the state enjoys large spatial correlations, implying that Dot1 overexpression also leads to patchy silencing, but unlike Dot1 inhibition, large patches of silencing domains established stochastically may be very long-lived in the cell. Such domains may be easier to establish and maintain in smaller chromatin regions, like HML/HMR, providing a possible explanation as to why Dot1 overexpression may have a weak effect at such loci. 

In the case of over-expression of Dot1 the system is driven out of the bistable region and is eventually pushed to a \emph{bivalent}-monostable regions. In the vein of the previous discussion, refer to Fig.~\ref{fig:corr-length2} and Fig~\ref{fig:rhobetacut2} once again. The range $\beta  \gtrsim 9 $ in Fig.~\ref{fig:corr-length2} and Fig.~\ref{fig:rhobetacut2} is currently relevant. In the wild type ($1.5 \lesssim \beta \lesssim 9$), we have strong correlations as expected. However, $\beta \gtrsim 9$ the system loses bistability and is \emph{bivalent}-monostable. The correlation length in this region is found to be comparable as in the bistable region (wild-type) well beyond the crossover. The strong persistence of correlations in spite of loss of stable silencing domain is a non-trivial prediction of our model. It suggests that for Dot1 overexpression silencing marks are established in patches that are long-lived and possibly heritable. A mechanistic explanation is: In the bistable region (wild-type), 
silencing domain is established and ambient Sir protein concentration is low. In the \emph{bivalent}-monostable region no such stable domain exists. However, increased effective rate of cooperativity in Sir recruitment ($\rho$) increases for increasing Dot1 activity ($\beta$) along the zero-velocity owing to depletion of chromatin-bound Sir proteins, see Eq.~\ref{eq:titration}. Therefore, Sir proteins that are chromatin bound stochastically form long-lived patches. 

\subsubsection*{Dot1 and Sas2 perturbations lead to distinct phases} 

It is now germane to contrast the effects of Dot1 perturbation and Sas2 perturbations in the light of robustness caused by multiple histone modifications. Both of the acetylation and methylation marks we consider are active marks. However, Sas2 inhibition pushes the wild-type system to regions of phase space where bistability loses robustness because the region of bistability narrows (the zero-velocity line runs towards a cusp), see Ref.~\cite{Sedighi2007,David-Rus2009} and Text S1. Dot1 inhibition pushes the system to regions where it continues to enjoy bistability, and the stable silenced state persists. However, a heritable pattern of silenced and active domains is compromised in Dot1 perturbations, as discussed above. Intuitively, the polymerization model of Sir recruitment which ensures continuous Sir occupancy also necessitates a histone modification feedback mechanism to counteract ectopic spread. The latter is, owing to the engineering limitations of local interactions and cell-cycle perturbations, 
easier to achieve using multiple histone marks~\cite{David-Rus2009}.

\section*{Discussion}

We have introduced and analyzed a model which captures the distinct mechanisms of cooperativity at the molecular level in the budding yeast silencing system; (a) cooperativity in the histone modification states, (b) cooperative interactions of the Sir proteins in a chromatin-bound-complex. Both of these mechanisms have been proposed by different groups, sometimes as mutually-exclusive ones~\cite{Millar2006, Turner2008, Latham2007, Kouzarides2007, Moazed2004, Rudner2005}. We show that these two mechanisms complement each other in designing epigenetic stability. Our model is minimal---all the interactions included are essential for establishment and inheritance of stable epigenetic fates. The model is build entirely upon known biochemical interactions. Though we explore design principles of the system using  inhibition/over-expression/knock-out of key proteins like Sir, Sas2 and Dot1, the model can be extended to include other silencing and active marks---and can inform, like it did for methylation, which of 
those marks are essential in resolving epigenetic fate, and which simply reinforces one fate over another.  

Our model makes the following predictions---
\begin{itemize}
\item{Dot1 over-expression and Sir2 inhibition can push the system to a novel \emph{bivalent} state exhibiting patchy silencing. We argue that the contradictory role of Dot1 in HML/HMR loci and telomeres can be reconciled by studying the correlations of Sir occupancy in the \emph{bivalent} state. Sir occupancy on the chromatin, instead of transcriptional activity, is a better reporter of the effects of Dot1 and Sir2 perturbations.}
\item{We compute Sir occupancy in simulations and compare its qualitative behavior to existing experimental work~\cite{Leeuwen2002a, Osborne2011, Rossmann2011, Takahashi2011}. Specifically, we observe that Dot1 inhibition can lead to higher fraction of chromatin-bound Sir and lower density of Sir in a loci owing to poor establishment of silencing domain, thereby causing Sir depletion. On the other hand, Dot1 overexpression can result in compact patches of Sir binding established stochastically. These patches are long-lived and may be heritable.}
\item{In comparing the behavior of Dot1 and Sas2 inhibition, we predict that while Dot1 inhibition does not eliminate stability of silenced domains, Sas2 does. We argue that methylation and acetylation play distinct roles in meeting the design requirements of bistability of states and establishment of inheritable domains of those states. } 
\end{itemize} 
Experiments focusing on direct measurement of Sir occupancy, preferably at single-cell resolution, as a function of tuning Dot1, Sas2 and Sir2's enzymatic activity will put to test several observations of the model---the emergence of the bivalent state, and the rich behavior of silencing domains under Sir titration. We have argued that inhibition/over-expression experiments can reveal key information about the engineering design of the system not observable in knock-out experiments. We hope that our model will sharpen experimental questions on the systems biology of budding yeast silencing. For the sake of brevity we have presented results for single perturbations; the model is especially informative for multiple perturbations, for example, double mutants. We are currently investigating such perturbations experimentally. 

After our paper was submitted, an experimental study by Grunstein lab~\cite{Kitada2012} has further revealed the role of H3K79 methylation in the silencing state of telomeres. Specifically, the transcription-mediated positive feedback of H3K79 methylation was discussed, and indications of our bivalent state where Sir occupancy and H3K79 methylation coexist were provided, though the response of this state to perturbations is still to be studied.

\section*{Materials and Methods}

The general model introduced in this paper is as follows. Define $\probP_i(S)$, $\probP_i(A)$, $\probP_i(M)$, $\probP_i(E)$ and $\probP_j(U)$ to be the probabilities of the nucleosome $i$ to be in one of the five mutually exclusive states, $S, A, M, E, U$ respectively. The following master equations define the time evolution of the probabilities (\emph{lattice model}), where summation over nucleosome index $j$ is for a suitable neighborhood of interactions with nucleosome $i$---
\begin{equation}
\begin{split} 
\label{eq:Model}
\frac{d \probP_i(S)}{dt} &= \left(\rho_0 + \rho \sum_{j\ne i}  \probP_j(S) \right)\, \probP_i(U) - \eta \, \probP_i(S), \\
\frac{d \probP_i(A)}{dt} &= \alpha\, \probP_i(U) - \Gamma \left(  \sum_{j\ne i} \probP_j(S) + \beta_0 +  \beta \sum_{j \ne i}  \probP_j(E)\right) \probP_i(A) - \eta \, \probP_i(A),\\
\frac{d \probP_i(M)}{dt} &= \left( \beta_0 +  \beta \sum_{j \ne i} \probP_j(E)  \right) \probP_i(U)  - \alpha \, \probP_i(M) - \eta \probP_i(M), \\
\frac{d \probP_i(E)}{dt} &= \left( \beta_0 +  \beta \sum_{j \ne i} \probP_j(E) \right) \probP_i(A) + \alpha \, \probP_i(M) - \eta \probP_i(E).
\end{split}
\end{equation}

We first analyze the model in the \emph{mean-field} limit. This implies that all spatial dependence---index $i$ and $j$ for the position along the chromosome---of all densities of marks are ignored and reactions occur in a well-mixed solution with infinite range of interaction. It is only in this limit that analytical solutions are obtainable. The mean-field analysis reveals all the phases of the model (\emph{phase space}) and guides the results of lattice simulations. where the lattice is the linear chromosome allowing reactions to occur only in a local neighborhood. Coexistence of silenced/active domains and stochastic spreading of silencing front can be explored in lattice simulations~\cite{Sedighi2007}. 

In the Text S1 under {\bf{Model I}}, we present the analytical mean-field solutions for the average densities of marks. All the phase diagrams presented are obtained from thse analytical solutions, which were numerically evaluated on a grid in the four dimensional model-parameter-space. The regions of bistability obtained in the mean-field picture typically shrink for the corresponding lattice model unless the range of cooperative interactions on the lattice is also infinite. Stochastic transition rates between silenced/active domains on the lattice reflect the timescales of stable establishment and heritability of such domains through cell-cycle perturbations. Therefore, we have computed the \emph{correlation length} (typical domain sizes) of silencing domains in lattice simulations.  

The construction of the model is minimal in the sense that two of the three cooperative terms---cooperative Sir binding and cooperative deacetylation---are essential for achieving bistability with respect to active and silent states. Cooperative Dot1 binding is non-essential, however, in Text S1 we show that establishment of well-defined silenced and active regions is compromised for the model with cooperative Dot1 rate $\beta=0$ and tunable basal Dot1 rate $\beta_0$. Intuitively, this is because bistability is achieved only for moderate to high basal Dot1 rate $\beta_0$, which makes the system fragile to loss of silencing marks (from cell-cycle perturbations) and domains of silencing are disrupted easily by basal Dot1 activity. In the above sense, the model construction is minimal. 

Though we have presented a specific model for concreteness, many the qualitative features, which includes the number of states and bistabilities supported, remain unchanged under certain plausible additional interactions: a phenomena known as the equivalence of dynamical systems~\cite{Kuznetsov1998}. More precisely, the region of bistability and the \emph{cusp bifurcation}~\cite{Kuznetsov1998} associated with it, remain even if the model is changed by introducing small perturbations. The \emph{transcritical bifurcations}~\cite{Kuznetsov1998} in our model; namely, the \emph{bivalent} state exchanging stability with the \emph{active} or the \emph{silenced} state, and the transition between the two types of bistable regions (see Fig.~\ref{fig:firstcut}); become sharp crossovers in the presence of small positive basal rates $\rho_0$ and $\beta_0$. We have verified that these basal rates need to be be small for robust bistability, see Text S1. Therefore, we expect the conclusions drawn from our simplified 
treatment to be applicable to a larger class of models. 

The simulations were performed using Gillespie algorithm on an one-dimensional lattice with $L$ lattice sites, with a fixed supply of Sir proteins $S_\text{tot} $ and a fixed volume of cell $V$. The specific choices for these parameters are quoted below. 

All rates are measured in units of the constant rate of loss of all marks---this loss rate models all cell-cycle perturbations as a continuous loss. One can, alternatively, consider discrete times where cell division halves all marks stochastically\cite{David-Rus2009}. Our loss rate (set to one) is rather exigent given that all the other rates are in the range O(1)-O(10). The neighborhood of interaction is $N$ nucleosomes and is used to compute the local densities of marks. This interaction neighborhood reflects the polymer nature of chromatin where neighboring nucleosomes can come in physical contact often; $N$ sets the scale for all our correlation length computation. We only report the qualitative trends of correlation lengths which do not depend on the precise value of $N$. Neighborhood weighting is as follows: For any nucleosome $i$ a window of $N$ neighboring nucleosomes indexed by $j$ is assigned exponentially decaying weights $w_{ij}$ centered on $i$; $\frac{\exp( - |x_{ij}|/N)}{\sum_{j=1}^{N} \exp(-
|x_{ij}|/N)} $, where $|x_{ij}|$ is the absolute value of the distance between $i$ and $j$. Therefore, local densities of $x \in \{ A,M,E,U \}$ are $\rho_i(x) = \sum_{j=1}^N w_{ij} \theta_j(x)$ where $\theta_j(x) \in \{0, 1\}$ is the indicator variable for the mark. For the mark $S$, $\rho_i(S) = \frac{S_{\text{ambient}}}{V} \sum_{j=1}^N w_{ij} \theta_j(S)$, where $S_\text{ambient} = S_\text{tot} - S_\text{bound}$, $S_\text{bound} = \sum_{i=1}^L \theta_i(S)$.

On the lattice the first ten nucleosomes (`telomeric end') are assumed to be in $S$ (\emph{silenced}) and the last ten to be in $E$ (\emph{active}). This choice simply sets the convention that in the bistable region, the silencing domain forms at the left end and active domain at the right end of the lattice. This boundary condition mimics the telomeric ends and the silencer regions of HML/HMR, which are `nucleation centers' of Sir proteins with high basal rate of Sir binding (i.e., high $\rho_0$). 

The system is allowed to reach steady state (with stochastically fluctuating but non-propagating front) and the effective $\rho(s_{\text{ambient}})$ determines the zero-velocity line. {\bf{Parameters:}} $ V = 50, L = 200, S_\text{tot} = 100, N=10$. In each iteration (\emph{time step}) a reaction is executed, where the reaction is chosen with probabilities proportional to their reaction rates. We perform ten million time steps for each choice of the tuned parameter. Initial time allowed to approach to steady state is  two hundred thousand steps. Samples are gathered every $10\times L$ time steps, thereby providing 5000 steady state samples to ascertain the effective $\rho(s_\text{ambient})$ against the tuned parameter (gray line in Figs.~\ref{fig:rhobetacut2} and Fig.~\ref{fig:zero-velo-Sir2}).

The correlation length of $S$ marks captures the typical sizes of silenced domains observed in the stochastic evolution of the system and can be computed using the Fourier Transform of the Green's function $\tilde{G}(k)$ for the $S$ marks, where $k$ is the Fourier space.  The Green's function is $G(k) = \tilde{\rho}(k) \tilde{\rho}(-k)$, where $\tilde{\rho}(k)$ is the Fourier transform of the local density $\rho(s)$ of $S$ marks where the local occupancy $s$ is a binary variable. We assume the standard form $\tilde{G}(k) = \frac{C_1}{k^2 + \xi^{-2}} + C_2$, where $\xi$ is the correlation length by definition and the $C$'s are fitting constants. $\tilde{G}(k)$ is computed by using a multi-taper (Slepian) estimate of the power spectrum using the Chronux toolbox~\cite{Bokil2010}. The larger fluctuation for larger correlation length in the computation is owing to the long-lived nature of larger domains. {{\bf Parameters:} $V = 150, L = 400, S_\text{tot} = 300, N=10, \rho_0 = 0.01, \beta_0 = 0.01$. Steady-state 
samples (5000) were generated from twenty million time steps sampled every $L \times 10$, for each value of $\beta \in [0,20]$ and $\Gamma \in [0,10]$ at increments of $0.5$. We utilize a nonlinear fitting tool in MATLAB~\textsuperscript{\textregistered} to fit the estimated power spectrum and compute correlation length $\xi$. 

\section*{Acknowledgments}
SM thanks Vijayalakshmi H. Nagaraj and Bruce W. Stillman's lab at Cold Spring Harbor Laboratory for useful discussion, and Justin B. Kinney for critical input on the manuscript.  


\section*{Figure Legends}

\begin{figure}[!ht]
\centering
\includegraphics[width=15cm]{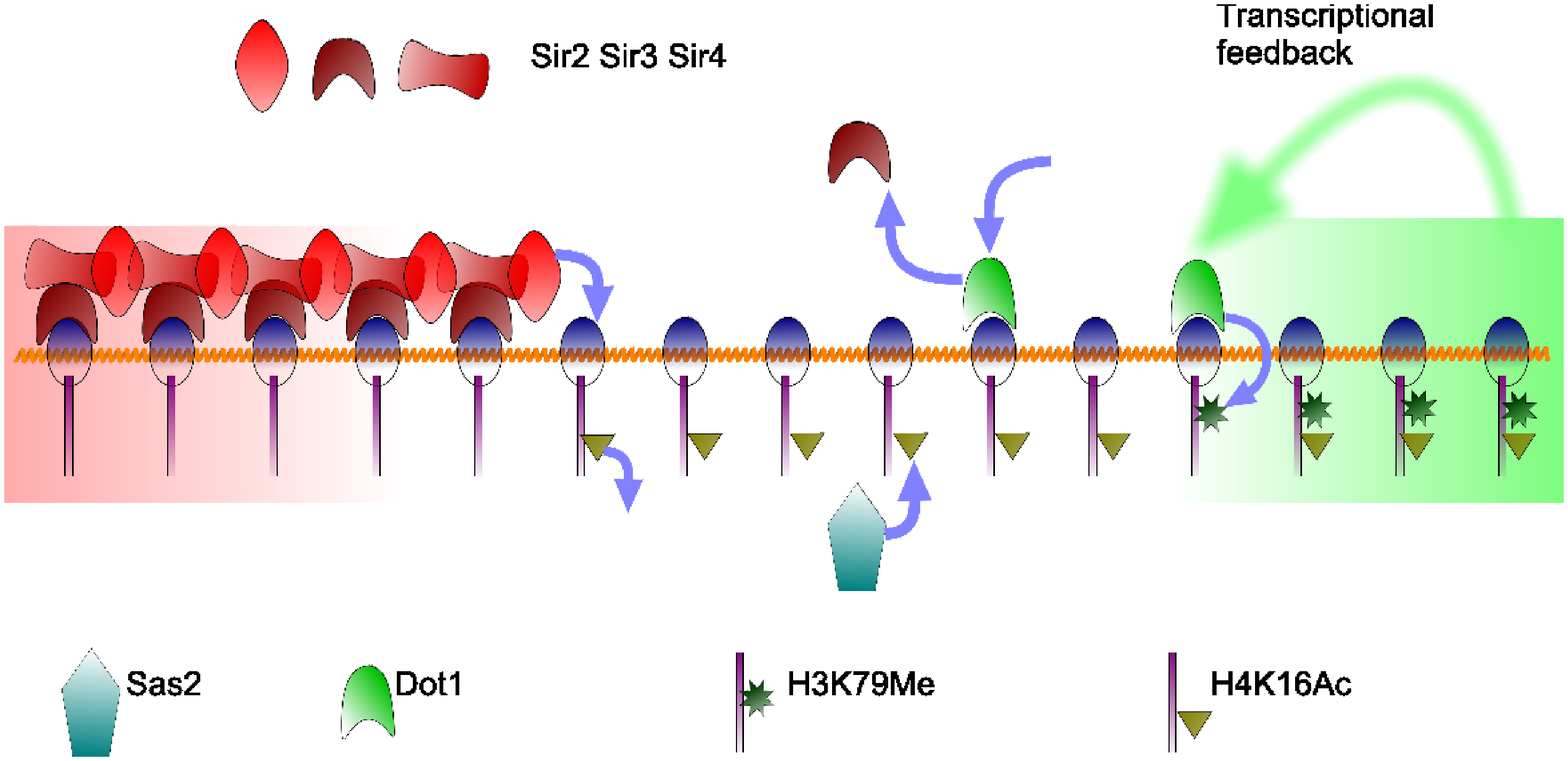}
\caption{The mechanisms of budding yeast silencing modelled: The red represent silenced and the green transcriptionally active region respectively. The transcriptional feedback is discussed in the text. We argue that other histone modifications perhaps reinforce, as opposed to drive, the distinct epigenetic fates.}
\label{fig:cartoon}
\end{figure}

\begin{figure}[!ht]
\centering
\includegraphics[width=15cm]{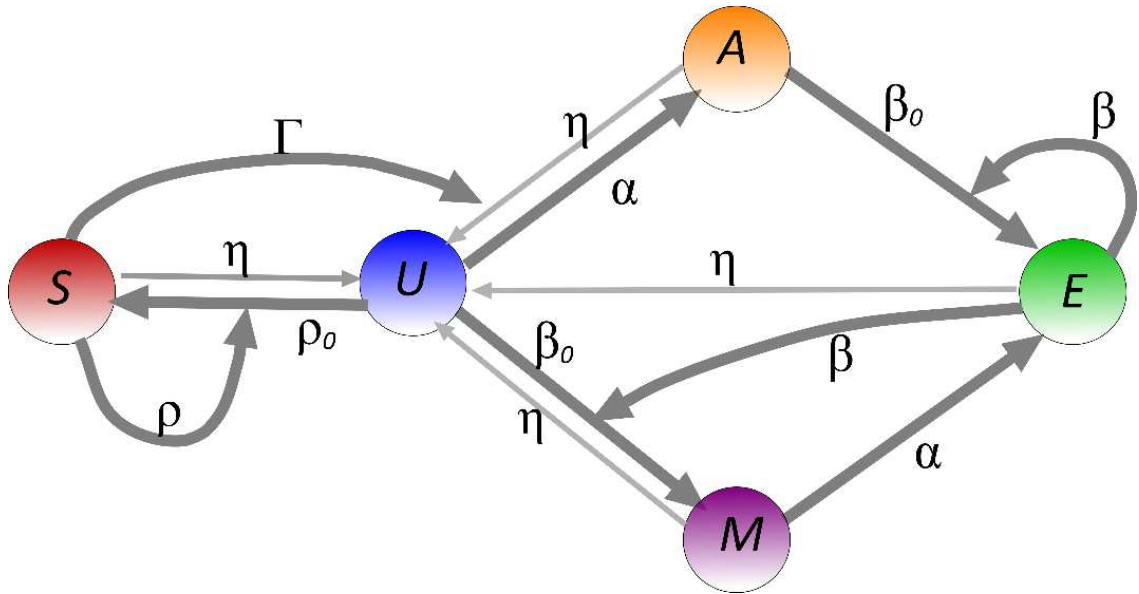}
\caption{The schematics of the minimal model: On a particular nucleosome, $S$ denotes the silencing mark, $U$ is no-mark, $A$ is the acetylation mark, $M$ is the methylation mark and $E$ is the active mark. All straight arrows represent local transitions of nucleosomal marks with rates shown. The thin straight arrows are the (low) overall rate of loss of histone marks. The curved arrows are cooperative interactions, where a neighboring nucleosome, bearing a mark (corresponding to the arrow's origin) influences the local transition of the nucleosome to another mark (corresponding to the arrow's target).}
\label{fig:cartoon-2} 
\end{figure}

\begin{figure}[!ht]
\centering
\includegraphics[width=15cm]{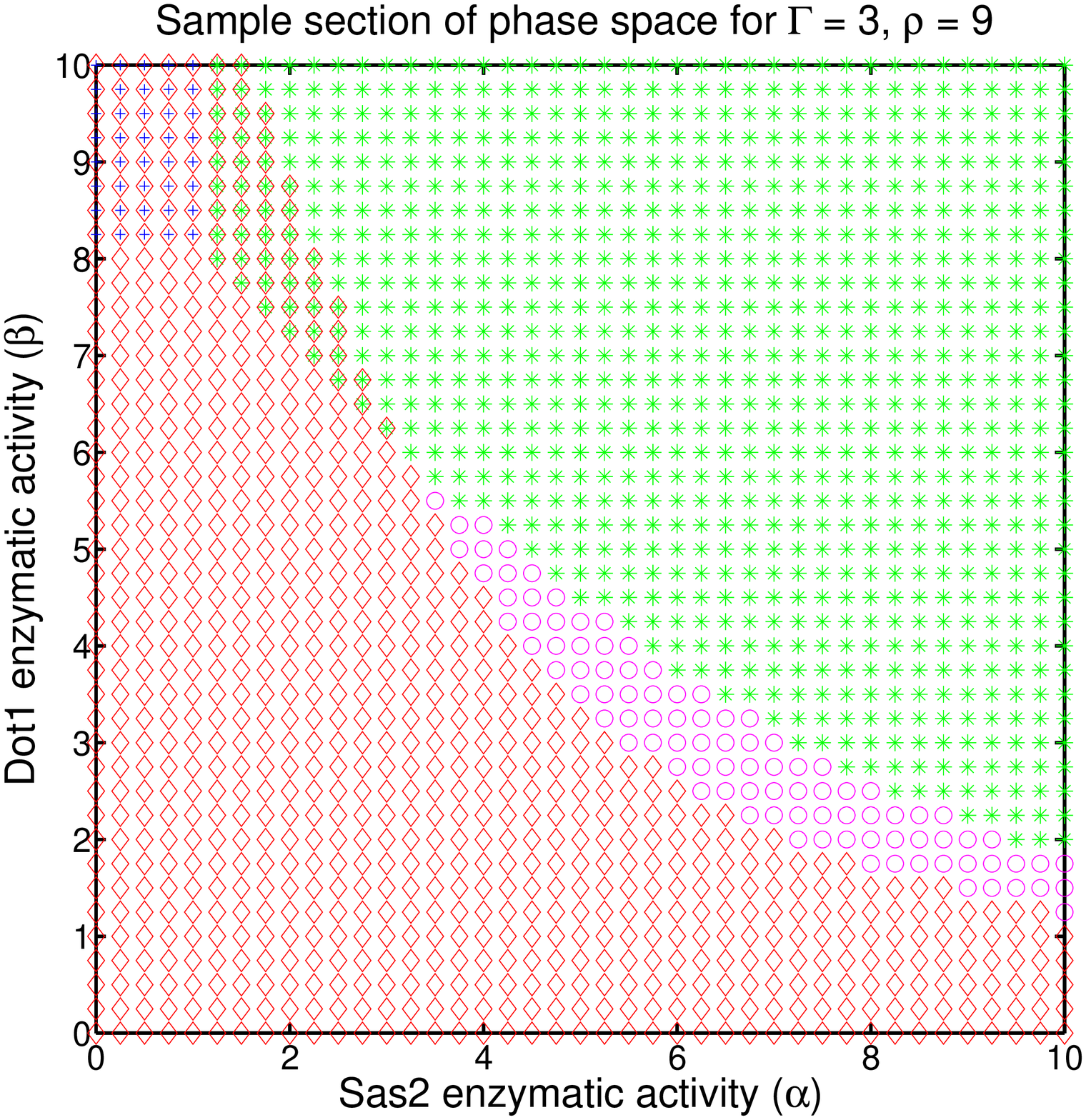}
\caption{Example section of the models phase diagram showing all the possible phases: The phase space is four dimensional. This section corresponds to fixing Sir2 deacetylase activity ($\Gamma$) and Sir cooperative binding rates $(\rho)$ and varying the Sas2 activity ($\alpha$) and cooperative Dot1 activity ($\beta$). Red diamonds represent region where \emph{silenced} state is stable, green stars where \emph{active} state is stable, magenta circles where \emph{bivalent} state is stable, blue crosses where \emph{intermediate} state is stable, and regions of overlap of symbols are bistable regions. The bistable solutions merge on $\alpha = 0$ line to a single solution. }
\label{fig:firstcut} 
\end{figure}

\begin{figure}[!ht]
	\begin{center}
 \includegraphics[width=15cm, scale=1]{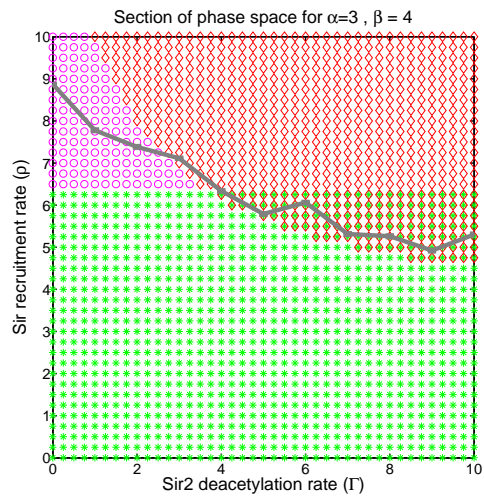}
	\caption{The zero velocity line: Section of phase diagram for varying Sir2 deacetylation activity ($\Gamma$) and Sir recruitment ($\rho$), and fixed Sas2 activity ( $\alpha = 3$) and Dot1 activity ($\beta = 4$).  For limited supply of silencer proteins the system self-adjusts the parameter $\rho(s_{\text{ambient}})$ for changing $\Gamma$ to settle on the zero velocity line (determined from simulations) drawn in gray.}
	\label{fig:zero-velo-Sir2}
	\end{center}
\end{figure}

\begin{figure}[!ht]
	\begin{center}
	\includegraphics[width=15cm, scale=1]{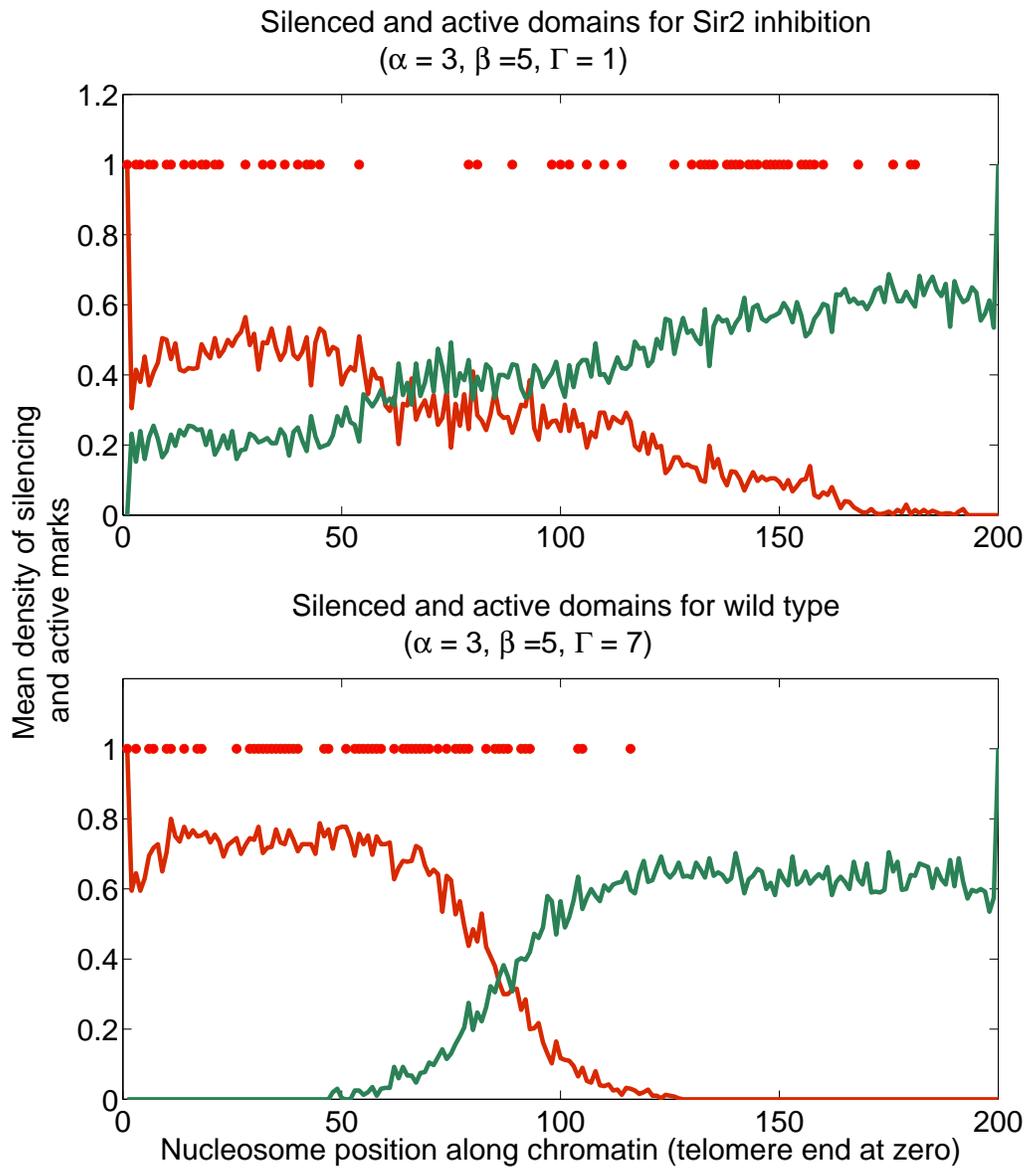}
	\caption{Effect of inhibiting Sir2 activity on Sir domains: Density profile of silencer mark S (red line) and active mark E (green line) in the \emph{bivalent} state encountered as a result of reducing inhibiting Sir2 activity ($\Gamma$), determined from lattice simulations. {\bf Top panel} shows disruption of silencing boundary for low Sir2 activity compared to ({\bf bottom panel}) wild-type choice of Sir2 activity. The density profiles are average over many stationary configurations. A single such configuration is shown in dots to emphasize the ill-defined Sir domain in individual samples.}
	\label{fig:Sas2MutantSpreading}
	\end{center}
\end{figure}

\begin{figure}[!ht]
	\begin{center}
	\includegraphics[width=15cm, scale=1]{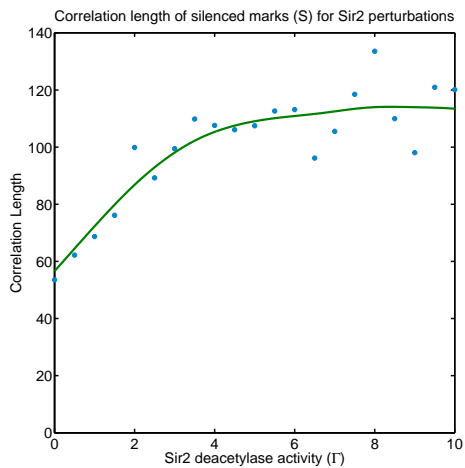}
	\caption{Effect of Sir2 perturbations: Correlation length of silenced mark ($S$) state of the system along the zero-velocity-line in Fig.~\ref{fig:zero-velo-Sir2}. Sir inhibition pushes the system to the bivalent-stable phase. The transition from active-silenced-bistable state (wild-type) to the bivalent-stable phase is approximately at $\Gamma \approx 4$. The correlation length is high, as expected, in the bistable region where silencing domains are established, however, the system continues to enjoy relatively strong correlation lengths in the bivalent-stable phase. The scale on the y-axis depends on the (unknown) biophysical parameters of the
wild-type system, we are only reporting the trend.}
	\label{fig:corr-length1}
	\end{center}
\end{figure}

\begin{figure}[!ht]
	\begin{center}
		\includegraphics[width=15cm, scale=1]{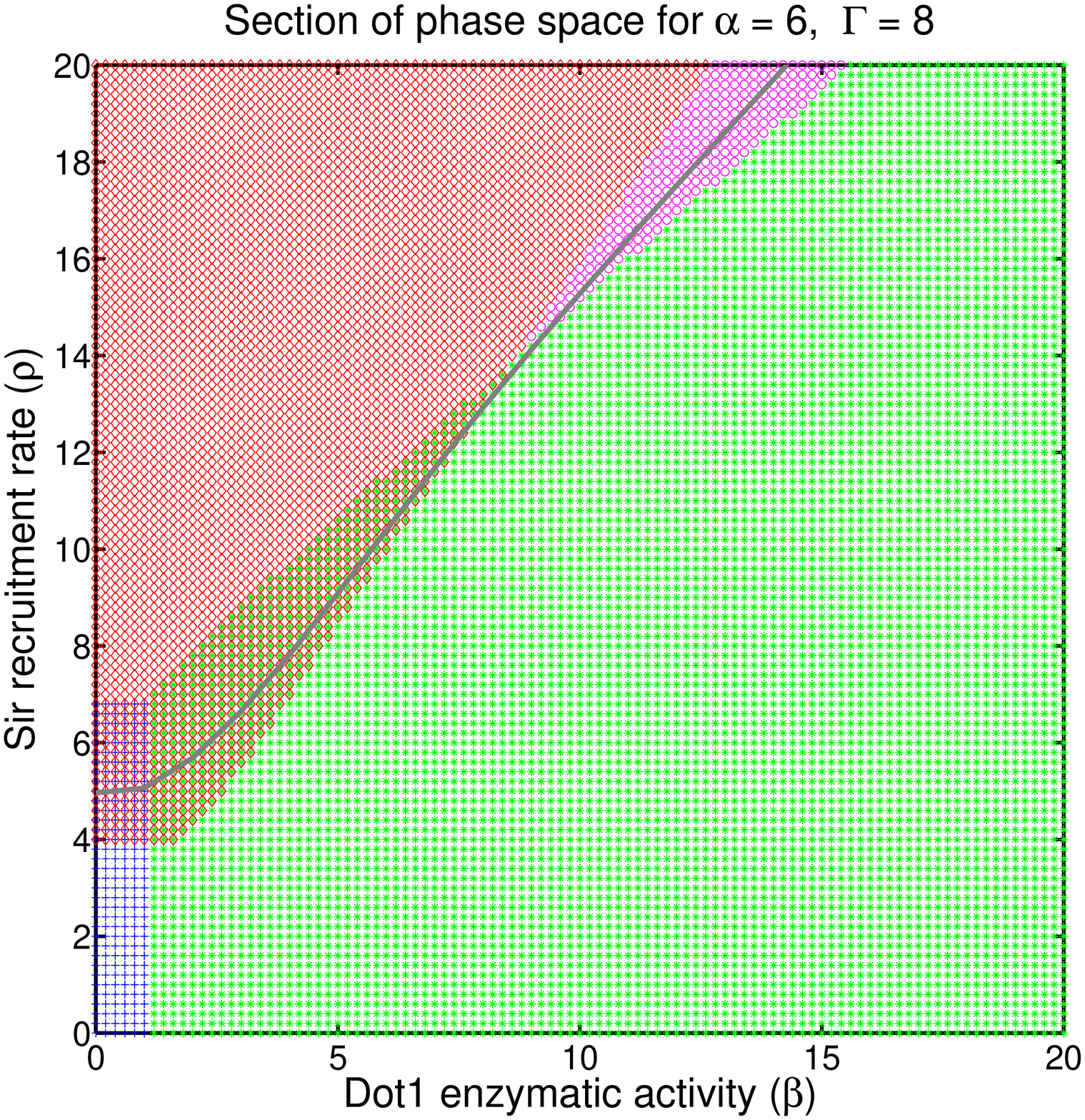}
	\caption{Inhibition and overexpression of Dot1: Section of phase space for fixed Sas2 activity ($\alpha$) and Sir2 activity ($\Gamma$) showing the effect of varying Dot1 activity ($\beta$) with Sir cooperative binding rate ($\rho$). The self-adjusting Sir binding owing to titration effect maintains the system on the zero-velocity line (grey line). The bivalent state (magenta) is stable for high $\beta$ and $\rho$---overexpression of Dot1 pushes the system into this phase. The wild-type bistability (silenced-active-bistable) region is the overlap of the green and red regions. Inhibition of Dot1 pushes the system to another bistable region with silenced and intermediate states stable---the overlap of the red and blue region.}
	\label{fig:rhobetacut2}
	\end{center}
\end{figure}

\begin{figure}[!ht]
	\begin{center}
		\includegraphics[width=15cm, scale=1]{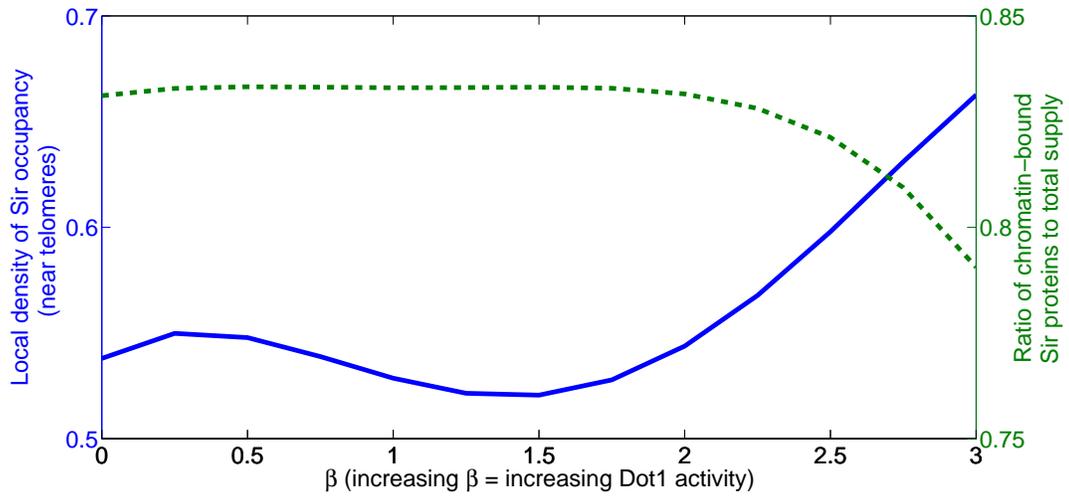}
	\caption{Effect of Dot1 inhibition on local Sir density and net chromatin-bound Sir: The density of Sir occupancy, determined from simulations, in telomeres (blue solid line) and fraction of Sir chromatin-bound (green dashed line) as a function of increasing Dot1 activity ($\beta$) starting from \emph{dot1$\varDelta$} upto wild-type values. The system is bistable for all values $\beta$ explored here, however, the nature of the bistability transitions from \emph{intermediate-silenced-bistable} to \emph{active-silenced-bistable} at $\beta \approx 1.5$. Note that for decreasing $\beta$, the local density of Sir protein decreases though the net fraction of chromatin bound Sir protein, counter-intuitively, increases. This is because compact silenced region are lost owing to Dot1 inhibition.}
	\label{fig:Dot1inhibitionSir}
	\end{center}
\end{figure} 

\begin{figure}[!ht]
\begin{center}

\includegraphics[width=15cm, scale=1]{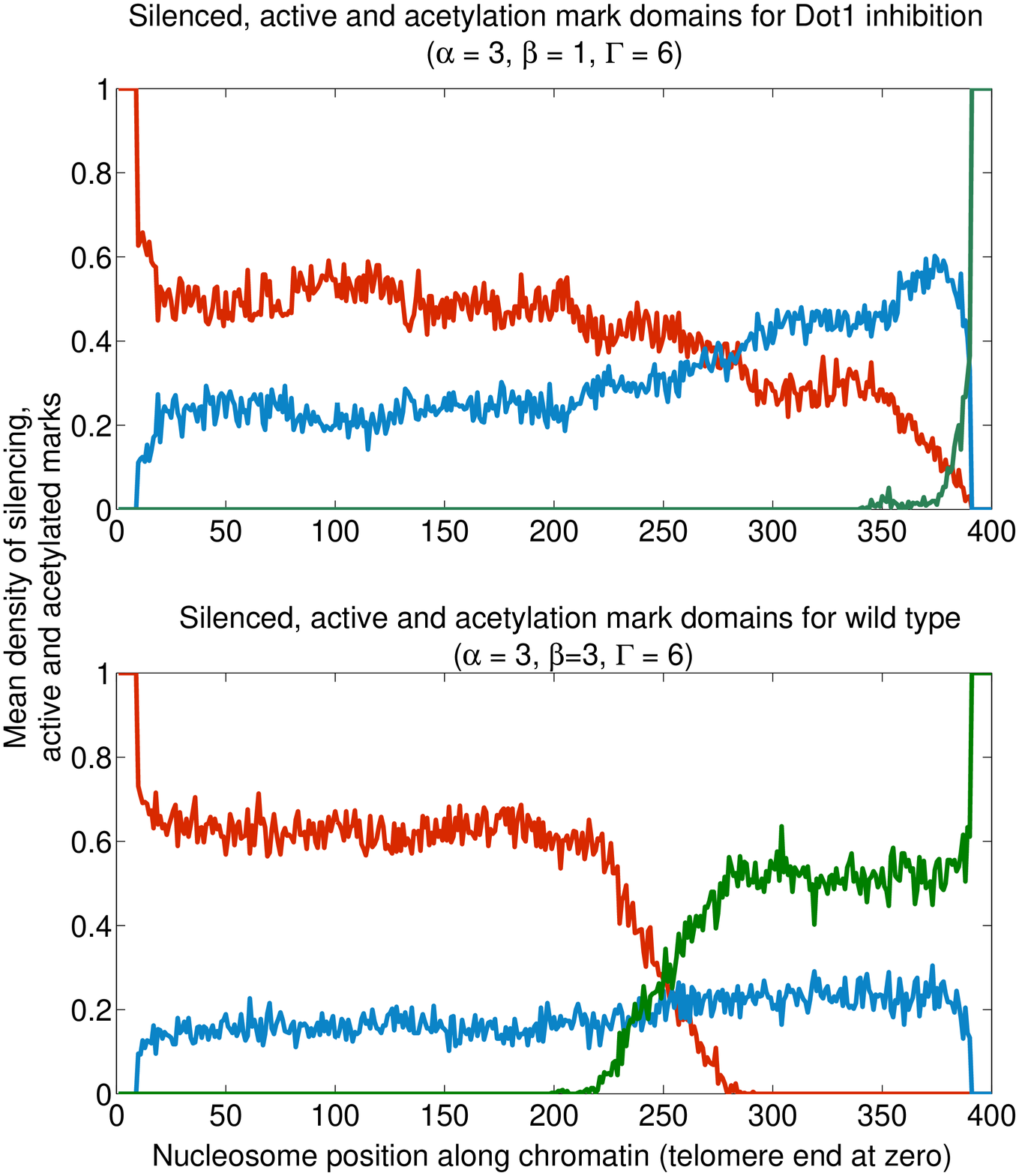}
\caption{Effect of Dot1 inhibition: Density profile of silenced mark $S$ (red line), active mark $E$  (green line) and acetylation mark $A$ (blue line) states where {\bf top panel} shows ill-defined establishment of silencing for Dot1 inhibition (low $\beta$) compared to wild-type values in the {\bf bottom panel} for which silencing domains are well-defined.}
\label{fig:Dot1boundaries}
\end{center}
\end{figure}

\begin{figure}[!ht]
	\begin{center}
	\includegraphics[width=15cm, scale=1]{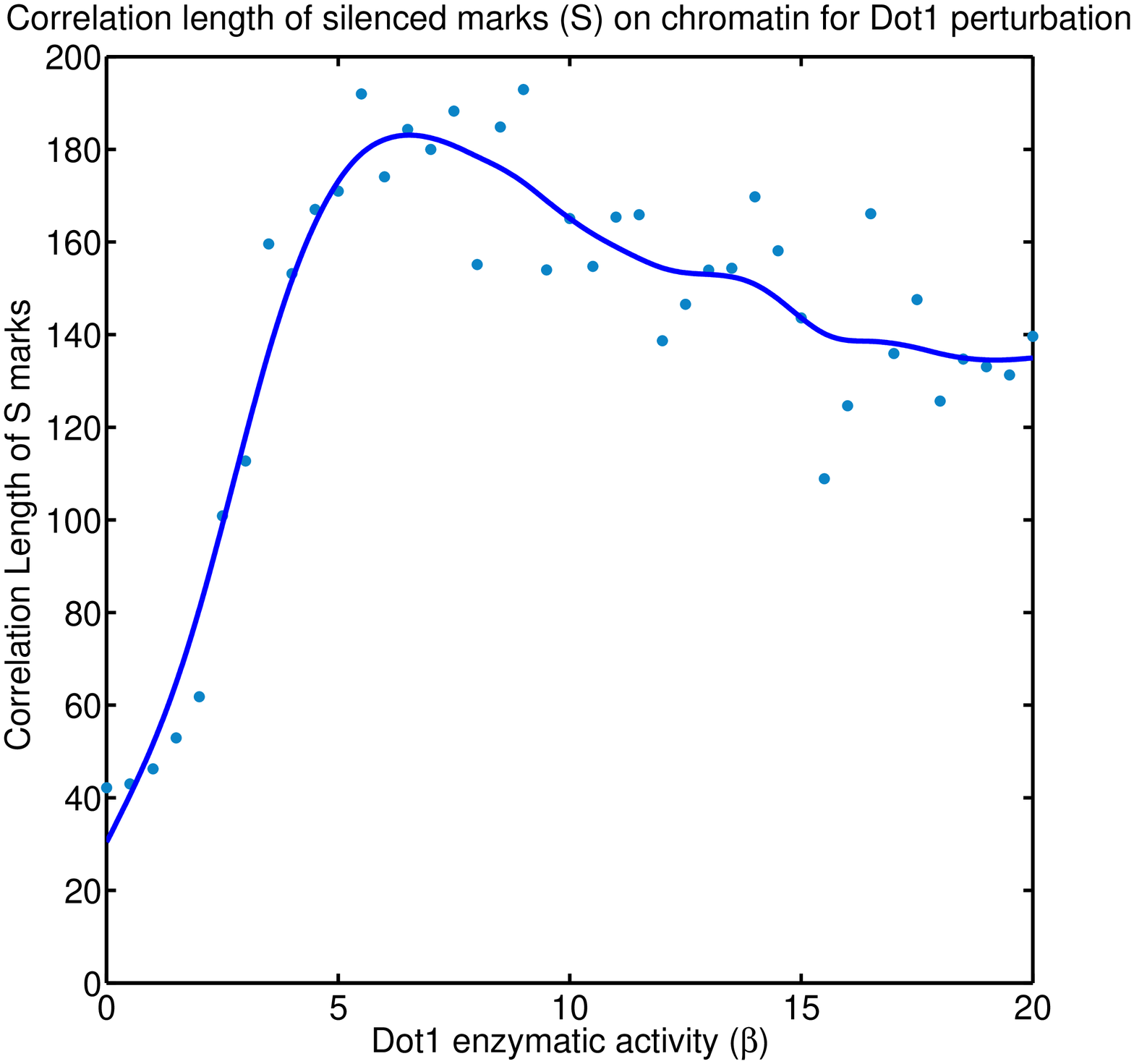}
	\caption{Effect of inhibition and overexpression of Dot1 activity on Sir domains: Correlation length of silenced mark ($S$) of the system along the zero-velocity-line in Fig.~\ref{fig:zero-velo-Sir2}. There are two transitions, one from \emph{silenced-intermediate-bistable} to \emph{silenced-active-bistable} at $\beta \approx 1.5$ and another from  \emph{silenced-active-bistable} to \emph{bivalent-monostable} at $\beta \approx 9$, see Fig.~\ref{fig:rhobetacut2}. The correlation length is high, as expected, in the bistable region (wild-type) where silencing domains are well-established. However, the system continues to enjoys high correlation length in the bivalent phase reached by Dot1 overepxression---long-lived patches of Sir domains persist in this region. For severe Dot1 inhibition, the correlation length drops in the silenced-intermediate-bistable phase, demonstrating that in spite of this state being bistable, silencing domains are ill-established. The scale on the y-axis depends on the (unknown) biophysical parameters of the wild-type system, we are only reporting the trend.}
	\label{fig:corr-length2}
	\end{center}
\end{figure}

\subsection*{Text S1} 
We have relegated the mathematical details of the model in Text S1. Analytical solution of the model in the mean-field limit is presented therein, and the nature of the different phases is summarized. We also compare related models with fewer or more biochemical interactions (parameters) in order to establish that the interactions we have considered are minimal for the engineering design criteria of epigenetic silencing. We also elaborate on Sir perturbations and Sas2 perturbations and discuss the nature of bifurcations/crossovers observed in the phase space.   


\end{document}